\documentclass{PoS}
\usepackage{amsmath}
\newcommand{\U}{\mathbf{U}}
\newcommand{\ud}{{\rm{d}}}
\newcommand{\Z}{\mathbb{Z}}
\newcommand{\redchisq}{\chi^2_{\tiny\mbox{r}}}

\title{Squared width and profile of the confining flux tube in the U(1) LGT in 3D}

\ShortTitle{Squared width and profile of the confining flux tube in the U(1) LGT in 3D}

\author{\speaker{Davide Vadacchino}$^{a,b}$, Michele~Caselle,$^b$ and Marco~Panero$^b$\\
        \llap{$^a$}Physics Department, College of Science, Swansea University\\
        Singleton Park, Swansea SA2 8PP, United Kingdom\\
        \llap{$^b$}Department of Physics, University of Turin \& INFN, Turin\\
        Via Pietro Giuria 1, I-10125 Turin, Italy\\
        E-mail: \email{vadacchi@to.infn.it}, \email{caselle@to.infn.it}, \email{panero@to.infn.it}
        }

\abstract{The dual formulation of the compact U(1) lattice gauge theory in three spacetime dimensions allows to finely study the squared width and the profile of the confining flux tube on a wide range of physical interquark distances. The results obtained in Monte Carlo simulations are compared with the predictions of the effective bosonic-string model and with the dual superconductor model. While the former fails at describing the data from a quantitative point of view, the latter is in good agreement with it. An interpretation of these results is proposed in light of the particular features of the U(1) lattice gauge theory in 3D and a comparison with non-Abelian gauge theories in four spacetime dimensions is discussed.
         }

\FullConference{34th annual International Symposium on Lattice Field Theory\\
                 24-30 July 2016\\
                 University of Southampton, UK}

\begin{document}

\section{Introduction}

As is well known, the low-energy regime of confining theories is dominated by configurations in which flux lines connecting a charge-anticharge pair are squeezed into a thin flux tube~\cite{Bali:1994de, Bali:2000gf, Lucini:2012gg, Brandt:2016xsp}. Since the study of its shape is one of the ways to gain insight into the confining mechanism, in the past a large numerical effort has been devoted to its characterization, focusing in particular on the transverse profile and on the squared string width, both for a ``mesonic'' and for a ``baryonic'' setup.

There are two competing ways to describe those two observables. The first is effective string theory (EST), in which the flux tube is described by a one dimensional Nambu-Goto (NG) string that acquires a width as a result of quantum fluctuations. At leading order (LO) in an expansion around the long-string limit, the model reduces to a free Gaussian string: its width is predicted to be~\cite{Luscher:1980iy}
\begin{equation}
	w^2(R) = \frac{1}{2\pi s}\log{\frac{R}{R_0}},
	\label{eq:w2ESTLO}
\end{equation}
where $s$ should coincide with the string tension $\sigma$ characterizing the model, and $R_0$ is a low-energy parameter. The latter takes into account that this description is only valid down to some distance at the scale of $1/\sqrt{\sigma}$. In this framework, a prediction has been computed up to the next-to-next-to-leading order~\cite{Gliozzi:2010zt, Gliozzi:2010zv} both at zero and finite temperature, and, until very recently, all the numerical studies have confirmed this prediction for both cases, both in the Abelian~\cite{Caselle:1995fh, Zach:1997yz, Koma:2003gi, Panero:2004zq, Panero:2005iu, Amado:2013rja} and non-Abelian case~\cite{Bakry:2010zt}; as shown in our recent work~\cite{Caselle:2016mqu}, however, the $\U(1)$ model in $D=3$ is a notable exception. The second way is related to the proposal of describing the vacuum of a confining theory as a dual superconductor~\cite{Mandelstam:1974pi, 'tHooft:1979uj}. According to this picture, the classical width of the flux tube should not depend on the separation of the charges and its transverse profile should decay exponentially, with a decay constant equal to the London penetration depth of the medium.

The $\U(1)$ lattice gauge theory (LGT) in $D=3$ is a perfect framework to compare these two perspectives: its analytical treatment in the semiclassical approximation provides us with a deep understanding of how confinement works in this particular case, while its exact mapping to a dual spin model makes it very easy and fast to simulate.

\section{The $\U(1)$ LGT in $D=3$}

The $\U(1)$ LGT in $D=3$ is defined on a threedimensional square lattice $\Lambda$ of spacing $a$ with the standard Wilson action,
\begin{equation}
	S_W = \frac{1}{ae^2}~\sum_{x\in\Lambda}~\sum_{1 \leq \mu < \nu \leq 3 } \left[ 1 -\mathrm{Re}\,U_{\mu\nu} \left( x \right) \right], 
	\label{eq:wilsonactionU1}
\end{equation}
where $U_{\mu\nu}(x)$ is the standard plaquette and $e$ is the coupling. Its partition function reads
\begin{equation}
	Z = \int \prod_{x\in\Lambda}\,\prod_{1\leq \mu<\nu\leq3} \ud U_\mu(x) e^{-S_W}
	\label{eq:partitionfunctionU1}
\end{equation}
where $\ud U_\mu(x)$ is the Haar measure on the group manifold.

The periodicity of $S_W$ with respect to the plaquette complex phase is the most fertile feature of this model. Much like what happens in the XY model in $D=2$, at weak coupling ($\beta \gg 1$), the partition function splits into a part describing spin waves and another describing topological excitations. The latter can be shown to behave as magnetic monopoles in the presence of an electric current loop, whose condensation is responsible for a non-vanishing mass gap $m_0$ and for the area-law decay of expectation values of large loops,
\begin{equation}
	m_0 a = c_0\sqrt{8 \pi^2 \beta}e^{-\pi^2 v(0)\beta},\;\;\; \sigma a^2 \geq
\frac{c_{\sigma}}{\sqrt{2\pi^2\beta}} e^{-\pi^2 v(0)\beta},
	\label{eq:m0andsigmaTH}
\end{equation}
where $v(0) \simeq 0.2527 \dots $ is the zero-distance Coulomb potential in lattice units, and semiclassically $c_0=c_{\sigma}/8=1$; for further details, see ref.~\cite{Caselle:2014eka}.

From the numerical point of view, we define an exact duality mapping to a spin model defined on the dual lattice with integer elementary degrees of freedom ${}^\star s$ and partition function
\begin{equation}
Z = \sum_{ \{ {}^\star s \in \Z \} } \prod_{\mbox{\tiny{bonds}}} I_{|\ud {}^\star s|} (\beta),
\end{equation}
where $\ud$ denotes a difference at neighboring sites, and $I_\nu(x)$ is the modified Bessel function of the second kind of order $\nu$. 

In the dual formulation it is easy to incorporate sources of the gauge field straight into the partition function. Representing a charge-anticharge pair at distance $R$ by two Polyakov loops winding around the system in opposite directions, the corresponding two-point correlation function $\langle P(0) P^\dag(x) \rangle$ reads
\begin{equation}
	\langle P(0) P^\dag(x) \rangle = \frac{Z_R}{Z} = \frac{1}{Z} \sum_{ \{ {}^\star s \in \Z \} } \prod_{\mbox{\tiny{bonds}}} I_{|\ud {}^\star s + {^\star n}|} (\beta) 
	\label{eq:PPcorrfunction}
\end{equation}
where the non-dynamical variable ${}^\star n$ is non-vanishing on the links dual to a surface bounded by the loops.

\section{The squared width and the flux-tube transverse profile}

In this work we studied two related quantities: the squared width of the flux tube and the transverse profile of the flux tube. The latter can be obtained from direct lattice calculations via 
\begin{equation}
e_l (x_t) = \frac{\langle P^\star (R) P(0) E_l (x_t) \rangle}{\langle P^\star (R) P(0) \rangle} - \langle P^\star (R) P(0) E_l (x_t) \rangle = \frac{\langle\ud^\star l + ^\star n\rangle}{\sqrt{\beta}},
	\label{eq:el}
\end{equation} 
where $x_t$ is the transverse direction in the symmetry plane of the charges. The squared string width $w^2$ is then defined as 
\begin{equation}
	w^2 = \frac{\int \ud x_t ~x_t^2~ e_l(x_t)}{\int \ud x_t~e_l(x_t)}.
	\label{eq:w2definition}
\end{equation}

A summary of the numerical setup can be found in the first four columns of tab.~\ref{tab:w2-fitlog}. For each value of $\beta$ we explored a wide range of charge separations, going from $R\sim a/\sqrt{\sigma}$ to several times that quantity. For each value of $R/a$, we measured $e_l(x_t)$ for $x_t$ going from the axis connecting the charges to values of $x_t$ such that $e_l\sim0$. Once the transverse profile was obtained, the computation of $w^2(R)$ was performed with a naive discretization of eq.~\ref{eq:w2definition}. The farthest points in the transverse profile, for which $e_l\sim 0$, just contribute to the systematic error of $w^2$; we define our estimate of the string width by truncating the sums at values of $|x_t|$ where $w^2$ reaches a plateau. The values thus obtained are reported in tab.~\ref{tab:w2results}.

\begin{table}[ht]
\centering
\begin{tabular}{|c|c|c|c|c|}
\hline
& \multicolumn{4}{|c|}{$w^2/a^2$} \\
\hline
$R/a$  & $\beta=1.7$ & $\beta=2.0$ & $\beta=2.2$ & $\beta=2.4$\\
\hline
 $4$ &    --      & $8.59(21)$  & $14.26(2)$ &     --      \\
 $6$ & $4.00(10)$ & $10.46(28)$ & $19.1(5)$  &     --      \\
 $8$ & $4.32(6)$  & $12.20(24)$ & $21.4(5)$  &     --      \\
$10$ & $4.74(6)$  & $13.75(27)$ & $26.2(5)$  & $47.2(1.4)$ \\
$12$ & $4.98(6)$  & $14.72(31)$ & $27.0(5)$  & $59.2(3.3)$ \\
$14$ & $5.23(6)$  & $15.49(27)$ & $30.4(6)$  & $55.4(2.8)$ \\
$16$ & $5.43(6)$  & $16.52(31)$ & $32.1(6)$  & $67.2(3.2)$ \\
$18$ & $5.63(8)$  & $16.70(31)$ & $34.5(6)$  & $65.1(3.3)$ \\
$20$ & $5.78(8)$  & $16.92(31)$ & $36.2(7)$  & $64.6(3.3)$ \\
$22$ & $5.81(6)$  & $18.0(4)$   & $38.5(7)$  & $71.5(3.2)$ \\
$24$ &    --      &    --       &    --      & $76.6(3.5)$ \\
$26$ &    --      &    --       &    --      & $75.3(3.3)$ \\
$28$ &    --      &    --       &    --      & $76.7(3.4)$ \\
\hline
\end{tabular}
\caption{Squared width $w^2$ of the flux tube at different values of $\beta=1/(ae^2)$, as a function of the distance $R$ between the static sources.}
\label{tab:w2results}
\end{table}

\section{Results and conclusions}

\begin{figure}[t]
\centerline{\includegraphics[width=0.8\textwidth]{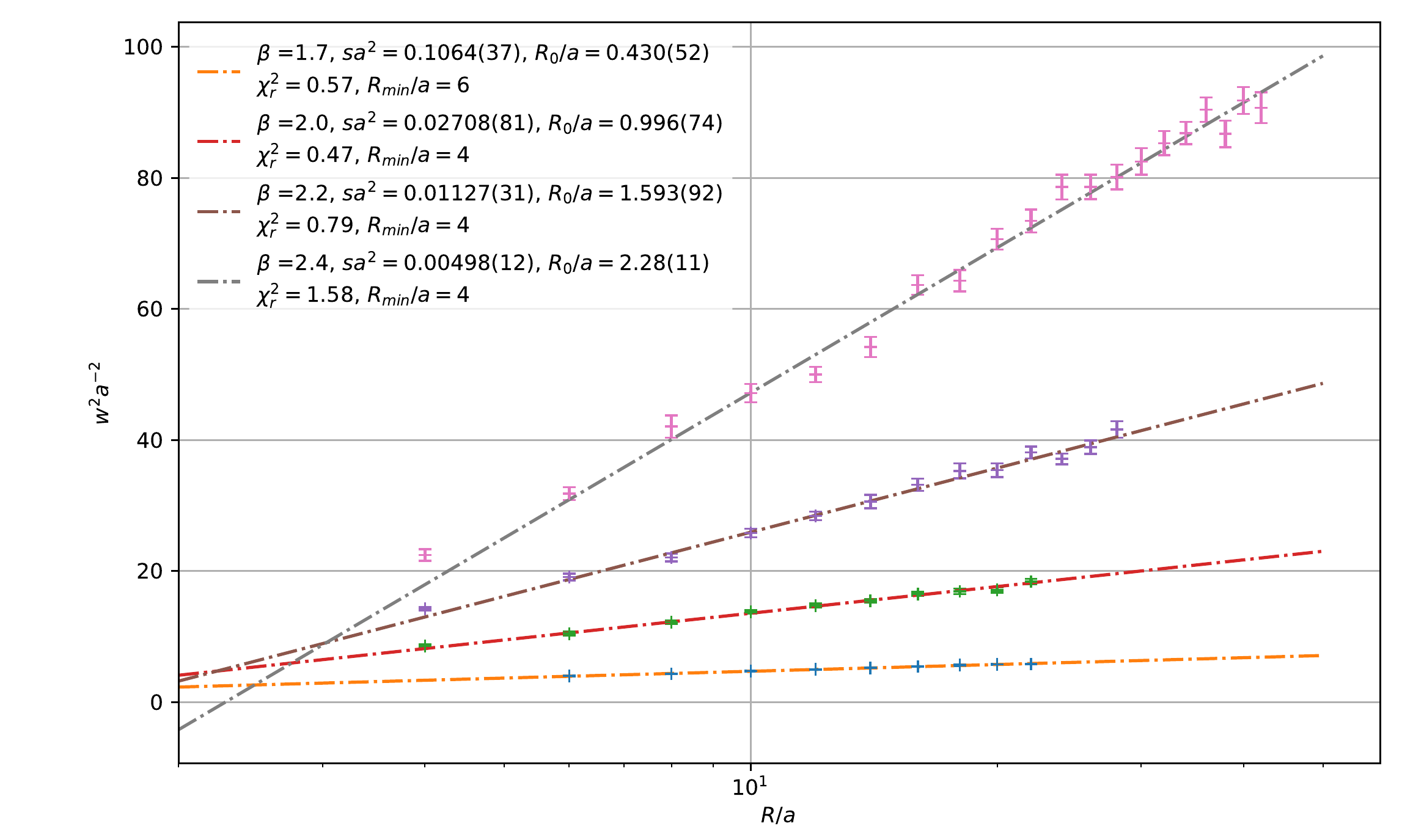}}
\caption{Results of the fits of the squared flux-tube width to the data, for the whole range of interquark distances.}
\label{fig:w2-fitlog}
\end{figure}

The first conclusion we can draw is that the string width is certainly not a constant with the charge separation. We must thus reject the dual superconductor approximation, at least at tree level and for finite lattice spacing. It is now natural to try and fit eq.~(\ref{eq:w2ESTLO}) to the data using $\sigma$ and $R_0$ as fitting parameters. We started by fitting the whole available $R/a$ range and then progressively discarded the smallest distances until an acceptable reduced $\chi^2$ (that we denote as $\redchisq$ in the following) was obtained. The results are reported in tab.~\ref{tab:w2-fitlog} and the fits are shown in fig.~\ref{fig:w2-fitlog}.

\begin{table}[!htpb]
\centering
\begin{tabular}{|c|c|c|c|c|c|c|c|c|}
\hline
$\beta$ &  $\sigma a^2$ & $m_0 a$ &  $L$, $N_t$ & $R_{\text{min}}/a$ & $s a^2$ & $R_0/a$ & $\redchisq$ &  d.o.f.\\
\hline
 $1.7$ & $ 0.122764(2)$  & $0.88(1)$ & $64$ & $6$ & $0.108(3)$    & $0.41(42)$  & $0.53$ & $7$ \\
 $2.0$ & $ 0.049364(2)$  & $0.44(1)$ & $64$ & $6$ & $0.0271(8)$   & $0.996(74)$  & $0.58$ & $8$ \\
 $2.2$ & $ 0.027322(2)$  & $0.27(1)$ & $96$ & $8$ & $0.0109(4)$   & $1.72(13)$ & $0.75$ & $9$ \\
 $2.4$ & $ 0.015456(7)$  & $0.165(10)$& $96$ & $12$ & $0.0044(2)$ & $3.01(25)$ & $0.48$ & $6$ \\
\hline
\end{tabular}
\caption{Results of the fits of $w^2$ to eq.~(\protect\ref{eq:w2ESTLO}), at the different values of $\beta$ and for the lattice sizes $L$, $N_t$ that we studied (for which the values of the string tension and lightest glueball mass are displayed in the second and third column). The fifth column shows the minimal value of $R/a$ included in the fit, while the sixth and the seventh column show the fitted parameters. The $\redchisq$ of the fit and the number of degrees of freedom (d.o.f.) are reported in the last two columns. Note that, although the dependence of the data on $R$ is logarithmic, as predicted by eq.~(\protect\ref{eq:w2ESTLO}), the fitted values of $s a^2$ are never compatible with $\sigma a^2$.
}
\label{tab:w2-fitlog}
\end{table}

As we can see, the LO order prediction from NG correctly describes the logarithmic dependence of $w^2$ on $R/a$, but the fitted value of the parameter $s$ is not the one predicted by EST. Note that for every value of $\beta$, the smallest value in the range of the fitted $R/a$ is always greater than $1/\sqrt{\sigma}$. Moreover, as we found by simple inspection, the best-fit values of the parameter $s$ actually scale with $m_0$ rather than with $\sigma$. 

The NLO correction to the logarithmic behavior, that we expect to give a detectable contribution to the fits in the short- to intermediate-distance range, does not improve the agreement between the data and the EST prediction. As fig.~\ref{fig:w2-fitlog} shows, the manifest logarithmic behavior persists down to distances comparable with $1/\sqrt{\sigma}$. 

\begin{figure}[!htpb]
\centerline{\includegraphics[width=0.5\textwidth]{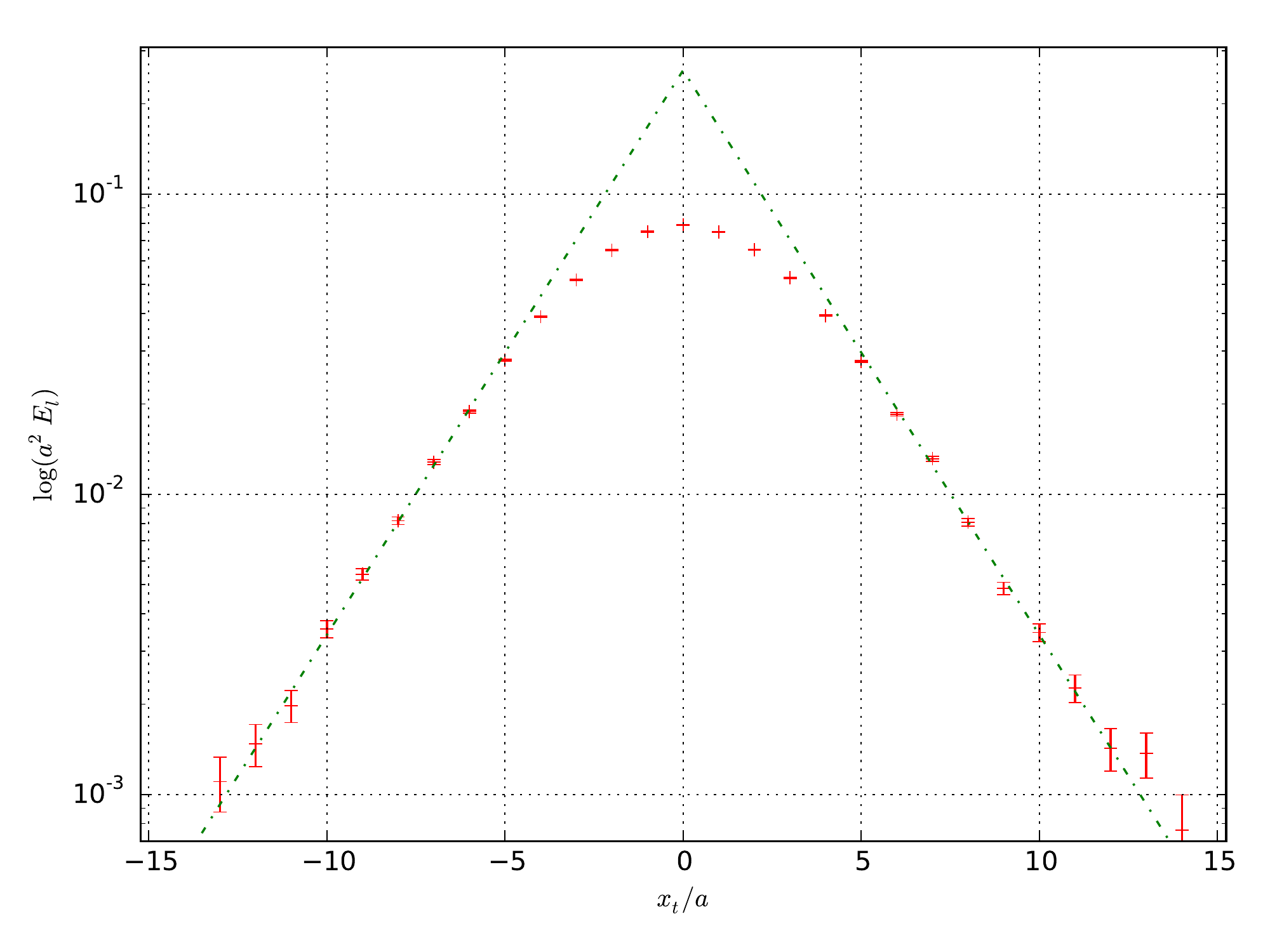}}
\caption{The transverse string profile, for $\beta=2.0$ and $R=16a$. Note that the vertical axis is displayed in a logarithmic scale. In this plot, the exponential decay of the tails of $e_l(x_t)$ manifests itself in the linear behavior observed for $\left| x_t \right| > 8a$: the green dash-dotted line is the result of the fit to eq.~(\protect\ref{eq:soldualsup}) in this range.}
\label{fig:logfitel}
\end{figure}

One can obtain further insight from the study of the transverse profile of the flux tube. 
EST predicts a Gaussian transverse profile $p(x_t) = C_0 \exp \left(-x_t^2 / \delta \right)$, while the solution of the equations of motion of the dual superconductor model yield an exponentially decaying profile, characterized by the London penetration depth $\lambda=1/m_v$,
\begin{equation}
	e_l(x_t) = \Phi ~ m_v^2 ~ e^{-m_v|x_t|}.
	\label{eq:soldualsup}
\end{equation}
Trying to fit any range of the profile with a Gaussian yields unacceptable $\redchisq$ values, while the tails, defined as the values of $e_l(x_t)$ with $x_t>x_t^c$, for some $x_t^c$ which is to be determined from the data, are correctly described by an exponential. For an example of fit of eq.~(\ref{eq:soldualsup} to the data see fig.~\ref{fig:logfitel} and tab.~\ref{tab:logfitel}. From the fitted values we see that, the quantity $m_v$ does not dependent on the separation of the charges and is compatible with $m_0$, as predicted by the dual-superconductor model. The quantities $x_t^c$ and $\Phi$ are weakly growing with $R$ for every value of the coupling constant. 

\begin{table}[!htpb]
\centering
\begin{tabular}{|c|c|c|c|c|c|c|c|}
\hline
$x_t^c/a$ & $\Phi$ & $a m_v$ & $\redchisq$ \\
\hline
$1 $ & $7.4(2)$   & $0.323(8)$ & $93.5$ \\
$2 $ & $6.9(1)$   & $0.368(6)$ & $21.5$ \\
$3 $ & $6.86(4)$  & $0.405(4)$ & $3.86$ \\
$4 $ & $6.99(3)$  & $0.428(4)$ & $1.48$ \\
$5 $ & $7.21(6)$  & $0.444(5)$ & $0.96$ \\
$6 $ & $7.25(13)$ & $0.447(7)$ & $0.96$ \\
$7 $ & $7.11(28)$ & $0.44(1)$  & $1.03$ \\
$8 $ & $7.3(6)$   & $0.44(2)$  & $1.06$ \\
$9 $ & $6.9(1.0)$ & $0.44(3)$  & $1.12$ \\
$10$ & $6.5(1.9)$ & $0.43(5)$  & $1.18$ \\
\hline 
\end{tabular}
\caption{Results of the fits of eq.~(\protect\ref{eq:soldualsup}) for the Monte~Carlo data obtained at $\beta=2.0$ and $R=10a$.}
\label{tab:logfitel}
\end{table}

The dual superconductor model is thus a good description for the tails of the transverse profile, but fails to describe a bell-shaped inner core of radius $x_t^c$.

As we already remarked, our study in ref.~\cite{Caselle:2016mqu} was motivated by the existence of two competing descriptions for the flux tube of a confining theory, and by the peculiar behavior of the $\U(1)$ model in $D=3$ in its flow towards the continuum limit. In this model it is possible to obtain very precise data for two observables related to the shape of the flux tube, that can be used to compare the agreement of both theories to the data. From the discussion above, we conclude that the prediction of EST up to NLO cannot correctly describe the data. Perhaps this is signaling us that, as was done for the static potential, the prediction should be supplemented by the contribution from a rigidity term in the effective string action~\cite{Caselle:2014eka}. We also showed that the flux-tube transverse profile is not Gaussian, and has exponentially decaying tails that can be characterized by $m_0$, as predicted by the dual-superconductor model. 

Note that in non-Abelian gauge theories, in which typically the $m_0/\sqrt{\sigma}$ ratio is roughly independent of $a$ (up to small discretization effects), the question whether an Abrikosov-like behavior of the flux tube is observed or not, cannot be answered by an analysis relying on the variation of the separation between the charges. It is then natural to focus on the transverse profile. Unfortunately, in the cases that we studied this observable gives no conclusive results, since it appears to include a mixture of Gaussian and exponential contributions.  

\acknowledgments
The work of D.~V. has been partially supported by the STFC grant ST/L000369/1.

\bibliographystyle{JHEP}
\bibliography{WidthRef}

\providecommand{\href}[2]{#2}\begingroup\raggedright\begin{thebibliography}{10}

\bibitem{Bali:1994de}
G.~S. Bali, K.~Schilling and C.~Schlichter, \emph{{Observing long color flux
  tubes in SU(2) lattice gauge theory}},
  \href{http://dx.doi.org/10.1103/PhysRevD.51.5165}{\emph{Phys. Rev.} {\bf D51}
  (1995) 5165--5198}, [\href{https://arxiv.org/abs/hep-lat/9409005}{{\tt
  hep-lat/9409005}}].

\bibitem{Bali:2000gf}
G.~S. Bali, \emph{{QCD forces and heavy quark bound states}},
  \href{http://dx.doi.org/10.1016/S0370-1573(00)00079-X}{\emph{Phys. Rept.}
  {\bf 343} (2001) 1--136}, [\href{https://arxiv.org/abs/hep-ph/0001312}{{\tt
  hep-ph/0001312}}].

\bibitem{Lucini:2012gg}
B.~Lucini and M.~Panero, \emph{{SU(N) gauge theories at large N}},
  \href{http://dx.doi.org/10.1016/j.physrep.2013.01.001}{\emph{Phys. Rept.}
  {\bf 526} (2013) 93--163}, [\href{https://arxiv.org/abs/1210.4997}{{\tt
  1210.4997}}].

\bibitem{Brandt:2016xsp}
B.~B. Brandt and M.~Meineri, \emph{{Effective string description of confining
  flux tubes}}, \href{http://dx.doi.org/10.1142/S0217751X16430016}{\emph{Int.
  J. Mod. Phys.} {\bf A31} (2016) 1643001},
  [\href{https://arxiv.org/abs/1603.06969}{{\tt 1603.06969}}].

\bibitem{Luscher:1980iy}
M.~L{\"u}scher, G.~M{\"u}nster and P.~Weisz, \emph{{How Thick Are
  Chromoelectric Flux Tubes?}},
  \href{http://dx.doi.org/10.1016/0550-3213(81)90151-6}{\emph{Nucl. Phys.} {\bf
  B180} (1981) 1}.

\bibitem{Gliozzi:2010zt}
F.~Gliozzi, M.~Pepe and U.-J. Wiese, \emph{{The Width of the Color Flux Tube at
  2-Loop Order}}, \href{http://dx.doi.org/10.1007/JHEP11(2010)053}{\emph{JHEP}
  {\bf 11} (2010) 053}, [\href{https://arxiv.org/abs/1006.2252}{{\tt
  1006.2252}}].

\bibitem{Gliozzi:2010zv}
F.~Gliozzi, M.~Pepe and U.-J. Wiese, \emph{{The Width of the Confining String
  in Yang-Mills Theory}},
  \href{http://dx.doi.org/10.1103/PhysRevLett.104.232001}{\emph{Phys. Rev.
  Lett.} {\bf 104} (2010) 232001}, [\href{https://arxiv.org/abs/1002.4888}{{\tt
  1002.4888}}].

\bibitem{Caselle:1995fh}
M.~Caselle, F.~Gliozzi, U.~Magnea and S.~Vinti, \emph{{Width of long color flux
  tubes in lattice gauge systems}},
  \href{http://dx.doi.org/10.1016/0550-3213(95)00639-7}{\emph{Nucl. Phys.} {\bf
  B460} (1996) 397--412}, [\href{https://arxiv.org/abs/hep-lat/9510019}{{\tt
  hep-lat/9510019}}].

\bibitem{Zach:1997yz}
M.~Zach, M.~Faber and P.~Skala, \emph{{Investigating confinement in dually
  transformed U(1) lattice gauge theory}},
  \href{http://dx.doi.org/10.1103/PhysRevD.57.123}{\emph{Phys. Rev.} {\bf D57}
  (1998) 123--131}, [\href{https://arxiv.org/abs/hep-lat/9705019}{{\tt
  hep-lat/9705019}}].

\bibitem{Koma:2003gi}
Y.~Koma, M.~Koma and P.~Majumdar, \emph{{Static potential, force, and flux tube
  profile in 4-D compact U(1) lattice gauge theory with the multilevel
  algorithm}},
  \href{http://dx.doi.org/10.1016/j.nuclphysb.2004.05.024}{\emph{Nucl. Phys.}
  {\bf B692} (2004) 209--231},
  [\href{https://arxiv.org/abs/hep-lat/0311016}{{\tt hep-lat/0311016}}].

\bibitem{Panero:2004zq}
M.~Panero, \emph{{A Numerical study of a confined Q anti-Q system in compact
  U(1) lattice gauge theory in 4D}},
  \href{http://dx.doi.org/10.1016/j.nuclphysbps.2004.11.203}{\emph{Nucl. Phys.
  Proc. Suppl.} {\bf 140} (2005) 665--667},
  [\href{https://arxiv.org/abs/hep-lat/0408002}{{\tt hep-lat/0408002}}].

\bibitem{Panero:2005iu}
M.~Panero, \emph{{A Numerical study of confinement in compact QED}},
  \href{http://dx.doi.org/10.1088/1126-6708/2005/05/066}{\emph{JHEP} {\bf 05}
  (2005) 066}, [\href{https://arxiv.org/abs/hep-lat/0503024}{{\tt
  hep-lat/0503024}}].

\bibitem{Amado:2013rja}
A.~Amado, N.~Cardoso and P.~Bicudo, \emph{{Flux tube widening in compact U (1)
  lattice gauge theory computed at $T < T_c$ with the multilevel method and
  GPUs}},  \href{https://arxiv.org/abs/1309.3859}{{\tt 1309.3859}}.

\bibitem{Bakry:2010zt}
A.~S. Bakry, D.~B. Leinweber, P.~J. Moran, A.~Sternbeck and A.~G. Williams,
  \emph{{String effects and the distribution of the glue in mesons at finite
  temperature}},
  \href{http://dx.doi.org/10.1103/PhysRevD.82.094503}{\emph{Phys. Rev.} {\bf
  D82} (2010) 094503}, [\href{https://arxiv.org/abs/1004.0782}{{\tt
  1004.0782}}].

\bibitem{Caselle:2016mqu}
M.~Caselle, M.~Panero and D.~Vadacchino, \emph{{Width of the flux tube in
  compact U(1) gauge theory in three dimensions}},
  \href{http://dx.doi.org/10.1007/JHEP02(2016)180}{\emph{JHEP} {\bf 1602}
  (2016) 180}, [\href{https://arxiv.org/abs/1601.07455}{{\tt 1601.07455}}].

\bibitem{Mandelstam:1974pi}
S.~Mandelstam, \emph{{Vortices and Quark Confinement in Nonabelian Gauge
  Theories}},
  \href{http://dx.doi.org/10.1016/0370-1573(76)90043-0}{\emph{Phys.Rept.} {\bf
  23} (1976) 245--249}.

\bibitem{'tHooft:1979uj}
G.~{'t Hooft}, \emph{{A Property of Electric and Magnetic Flux in Nonabelian
  Gauge Theories}},
  \href{http://dx.doi.org/10.1016/0550-3213(79)90595-9}{\emph{Nucl. Phys.} {\bf
  B153} (1979) 141}.

\bibitem{Caselle:2014eka}
M.~Caselle, M.~Panero, R.~Pellegrini and D.~Vadacchino, \emph{{A different kind
  of string}}, \href{http://dx.doi.org/10.1007/JHEP01(2015)105}{\emph{JHEP}
  {\bf 1501} (2015) 105}, [\href{https://arxiv.org/abs/1406.5127}{{\tt
  1406.5127}}].

\end{thebibliography}\endgroup

\end{document}